\documentclass[prl,twocolumn,superscriptaddress]{revtex4-1}

\usepackage{amssymb,amsmath,amsfonts}
\usepackage{microtype,graphicx,multirow,color,booktabs,braket,siunitx,float}

\renewcommand{\figurename}{Figure}
\renewcommand{\tablename}{Table}
\renewcommand{\tabcolsep}{4.5pt}
\renewcommand{\thetable}{\arabic{table}} 

\newsavebox{\foobox}
\newcommand{\slantbox}[2][0]{\mbox{%
        \sbox{\foobox}{#2}%
        \hskip\wd\foobox
        \pdfsave
        \pdfsetmatrix{1 0 #1 1}%
        \llap{\usebox{\foobox}}%
        \pdfrestore
}}
\newcommand\unslant[2][-.25]{\slantbox[#1]{$#2$}}
\newcommand{\upalpha}{\text{\unslant\alpha}}
\newcommand{\upbeta}{\text{\unslant\beta}}

\begin{document}

\title{Distinguishing the roles of energy funnelling and delocalization in photosynthetic\\ light harvesting}

\author{Sima Baghbanzadeh}
\affiliation{Department of Physics, Sharif University of Technology, Tehran, Iran}
\affiliation{Centre for Engineered Quantum Systems, Centre for Quantum Computation and Communication Technology, and School of Mathematics and Physics, The University of Queensland, Brisbane QLD 4072, Australia}
\author{Ivan Kassal}
\email{i.kassal@uq.edu.au}
\affiliation{Centre for Engineered Quantum Systems, Centre for Quantum Computation and Communication Technology, and School of Mathematics and Physics, The University of Queensland, Brisbane QLD 4072, Australia}

\begin{abstract}
Photosynthetic complexes improve the transfer of excitation energy from peripheral antennas to reaction centers in several ways. 
In particular, a downward energy funnel can direct excitons in the right direction, while coherent excitonic delocalization can enhance transfer rates through the cooperative phenomenon of supertransfer. 
However, isolating the role of purely coherent effects is difficult because any change to the delocalization also changes the energy landscape. 
Here, we show that the relative importance of the two processes can be determined by comparing the natural light-harvesting apparatus with counterfactual models in which the delocalization and the energy landscape are altered. 
Applied to the example of purple bacteria, our approach shows that although supertransfer does enhance the rates somewhat, the energetic funnelling plays the decisive role. 
Because delocalization has a minor role (and is sometimes detrimental), it is most likely not adaptive, being a side-effect of the dense chlorophyll packing that evolved to increase light absorption per reaction center.
\end{abstract}

\maketitle

Photosynthetic organisms harvest light using antenna complexes containing many chlorophyll molecules~\cite{Blankenship2014}. The energy collected by the antennas is then transmitted, through excitonic energy transfer (EET)~\cite{MayKuhn}, to a reaction center (RC), where it drives the first chemical reactions of photosynthesis. The thorough study of EET in photosynthetic antennas has been motivated, in part, by the prospect of learning how to design more efficient artificial light-harvesting devices~\cite{Scholes:2011uj,Laos:2014jf}. 

It has long been recognized that excitons in many photosynthetic complexes are directed toward the RC energetically: if the antennas lie higher in energy than the RC, the excitons can spontaneously funnel to the RC. A more recent discovery is that coherent mechanisms can also enhance light-harvesting efficiency. In particular, excitonic eigenstates may be localized or delocalized over a number of molecules, depending on the strength of their couplings~\cite{Leupold1996,Pullerits1996,Monshouwer1997,Oijen1999,MayKuhn}. Delocalization---i.e., coherence in the site basis---makes the aggregate behave differently than a single chlorophyll molecule, and phenomena such as superradiance~\cite{Dicke1954,Meier:1997fq}, superabsorption~\cite{Higgins2014}, and supertransfer~\cite{Strek1977,Scholes:2002ie,Lloyd:2010fz,Kassal2013} can occur in densely packed aggregates. Specifically, supertransfer occurs when delocalization in the donor and/or the acceptor enhances the rate of the (incoherent) EET between them. 

The presence of both funnelling and supertransfer suggests that their contributions to the efficiency could be quantified. However, the two effects are too closely related for such a separation to be easily carried out; in particular, a change to the extent of delocalization requires changing excitonic couplings, which also determine the energy landscape. In other words, because delocalization and the energy landscape are intimately connected, it is not sufficient to alter one property to see what happens to the efficiency, because doing so also alters the other property as well.

Here, we show that the roles of the two processes can be separated by constructing counterfactual light-harvesting complexes that reflect plausible evolutionary alternatives. In particular, screening thousands of complexes with varying energetic landscapes and extents of delocalization allows the effects of changing one property to be examined while keeping the other as constant as possible.

We describe this approach through its application to the strikingly symmetric antenna complexes of purple bacteria~\cite{Cogdell:2006ko}, which feature tightly packed bacteriochlorophylls and considerable excitonic delocalization~\cite{Hu1997,Scholes:1999tm,Damjanovic2000,Ritz2001,Sener2007,Olaya-Castro2008,Linnanto2009,Strumpfer2009,Sener:2010ct,Strumpfer2012a,Strumpfer2012b,Cleary:2013dl}. This delocalization is known to give rise to supertransfer, in particular for EET within the LH2 complex~\cite{Mukai1999,Scholes2000,Jang2004}.
In principle, supertransfer between different complexes should also occur, and although it has not been studied as well as intra-complex supertransfer, it has been proposed to explain the high light-harvesting efficiency~\cite{Strumpfer2012b,Lloyd:2010fz}.

We show that although supertransfer is present in purple bacteria, it is not essential for efficient light harvesting. For example, EET efficiency can remain roughly as high even if half of the pigments are removed, weakening the couplings and localizing the states. When delocalization is removed from only some of the complexes, the efficiency can change drastically, but these changes are almost entirely due to shifts in energy levels; even in the worst cases, modest modifications of the site energies can restore the high efficiency. Indeed, in the presence of a strong funnel, the efficiency can be high regardless of delocalization. This suggests that the evolutionary advantage of densely packed chlorophylls is that they enhance the absorption cross-section per RC, while the delocalization is merely a side-effect of the dense packing. In evolutionary language, delocalization is a spandrel, not an adaptation \cite{Gould:1997fv}.

\section{Model}

\subsection{Structure}

\begin{figure}[t]
    \centering
    \includegraphics[width=8cm]{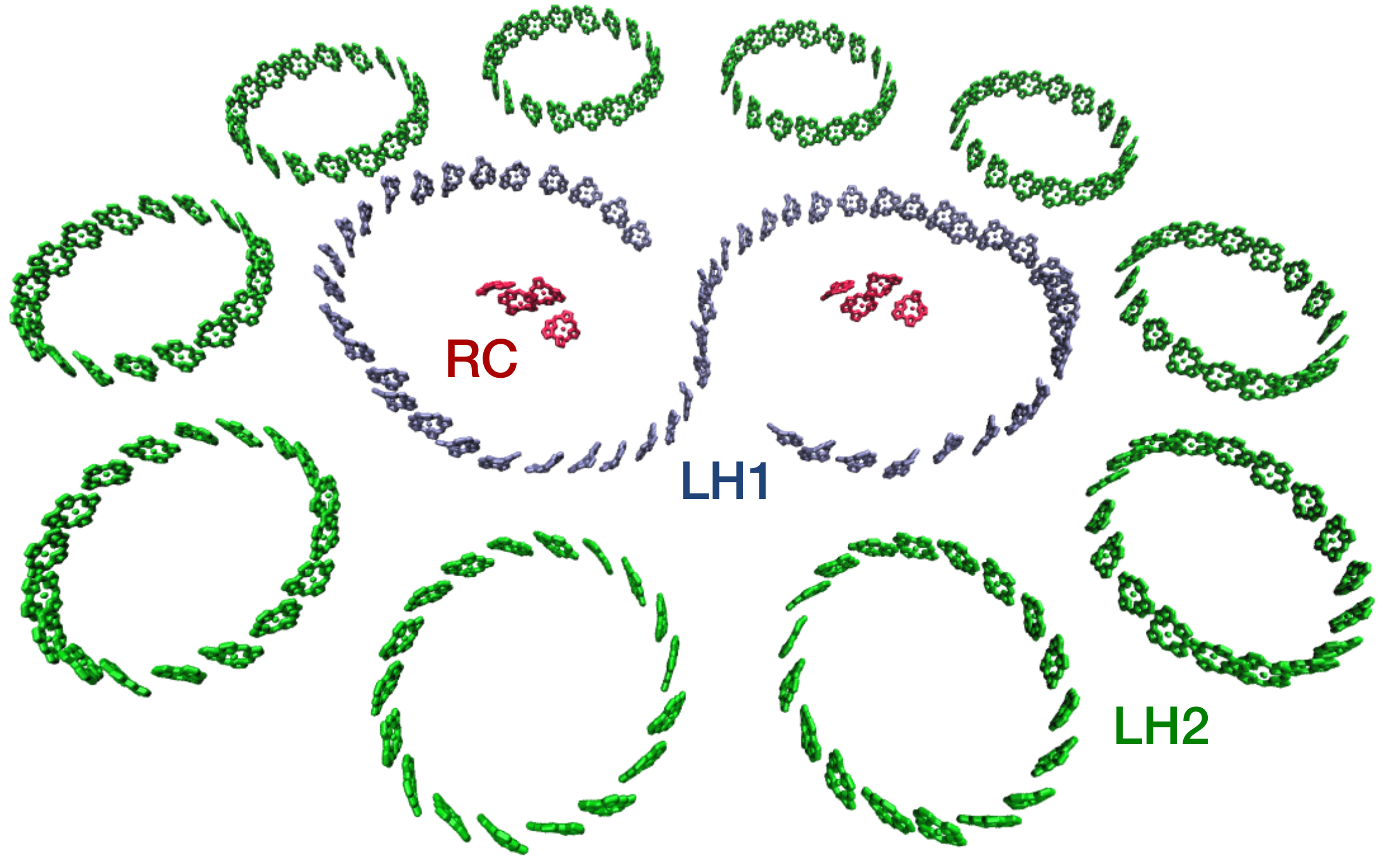}
    \caption{Model of the photosynthetic apparatus of \textit{Rh.\;sphaeroides}, including the reaction center surrounded by the antenna complexes LH1 and LH2 (only B850 subunit shown). Drawn using VMD~\cite{VMD}.}
    \label{fig:structure}
\end{figure}

We study the photosynthetic apparatus of the purple bacterium \textit{Rhodobacter sphaeroides}, which consists of reaction centers surrounded by two types of transmembrane antenna complexes, LH1 and LH2~\cite{Cogdell:2006ko}. Although all complexes can absorb light directly, the dominant energy-transfer pathway is LH2~$\to$~LH1~$\to$~RC. The complexes can arrange themselves in many ways \cite{Scheuring:2004cf,Scheuring:2009jb}; as a representative example we consider an array of ten evenly spaced LH2s surrounding the core LH1-RC complex, as shown in Fig.~\ref{fig:structure}.

The coordinates of the pigments in Fig.~\ref{fig:structure} are taken from the available crystal structures~\cite{Papiz2003,Qian2013}. Each LH1 complex consists of 56 bacteriochlorophyll \textit{a} molecules (BChl) with an average nearest-neighbor Mg--Mg distance of \SI{8.5}{\AA} and immediately encircling two RCs~\cite{Qian2013}. These BChls absorb at \SI{875}{nm} and are thus referred to as the B875 aggregate. Each RC comprizes four BChls, two in the tightly-coupled special pair and two accessory ones. The absorption spectrum of the RC has peaks at \SI{865}{nm} and \SI{802}{nm}, corresponding to the special pair and the accessory BChls, respectively.
  
LH2 complexes surrounding the LH1-RC core include two rings each, one with 9 BChls (the B800 subunit) and the other 18 (B850)~\cite{Papiz2003}. We only consider the B850 ring because B800 excitons are transferred to B850 efficiently and quickly (\SI{700}{fs}~\cite{Shreve1991}) and efficiently using supertransfer~\cite{Mukai1999,Scholes2000,Jang2004}, making their ultimate fate nearly identical to that of excitons starting at B850. The average \mbox{Mg--Mg} distance between BChls in B850 is \SI{9.0}{\AA}.

The distances between complexes have been determined by atomic force microscopy (AFM) \cite{Scheuring:2004cf,Scheuring:2009jb}, and our model uses the most common center-to-center spacing between LH2s, \SI{75}{\AA}. This gives a distance of \SI{22.5}{\AA} between nearest Mg atoms in different LH2s, which we also used to set the separation between the LH2s and LH1. 

\begin{figure*}[t]
    \centering
    \includegraphics{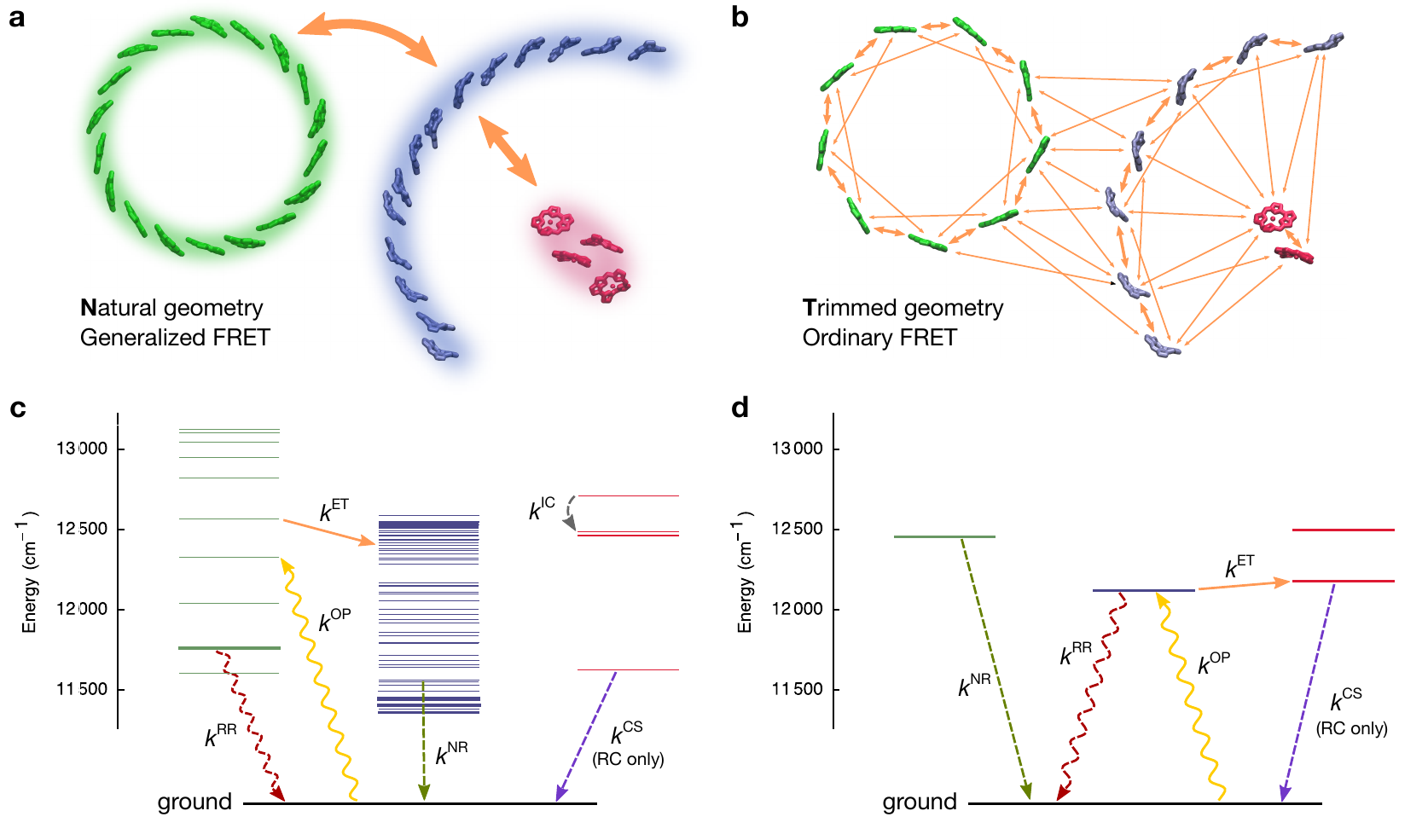}
    \caption{Details of the model.
    \textbf{(a)} In the natural geometry (N), all the pigments are included. Due to strong intra-complex couplings, excitations are delocalized and inter-complex energy transfer is described by generalized FRET, which transfers population from donor eigenstates to acceptor eigenstates.
    \textbf{(b)} In the trimmed geometry (T), every second pigment is removed to weaken the couplings and localize the excitons, so that energy transfer is described by ordinary, site-to-site FRET. Only the larger FRET rates are shown.
    \textbf{(c)} Excitonic energy levels in the natural geometry ($S$ parameter set). Brighter states are drawn thicker. Examples of each process included in our model are shown: energy transfer ($k^\mathrm{ET}$), optical pumping ($k^\mathrm{OP}$), radiative relaxation ($k^\mathrm{RR}$), non-radiative relaxation ($k^\mathrm{NR}$), internal conversion ($k^\mathrm{IC}$), and charge separation in the RC ($k^\mathrm{CS}$).
    \textbf{(d)} In the trimmed geometry, the site energies are not split by the excitonic couplings. The energy levels in the $R$ parameter set are shown in Supplementary Fig. 1.
    }
    \label{fig:rates}
\end{figure*}

We refer to the geometry just described as the natural geometry (N), to distinguish it from the trimmed geometry (T) (Fig.~\ref{fig:rates}b), in which every other BChl was removed to reduce nearest-neighbor couplings and encourage exciton localization (see below). For example, the distance between BChls in B850 roughly doubles upon trimming, becoming comparable to that in the B800 ring, where delocalization is known to be minor. To trim each RC, out of the four BChls, we kept one of the special pair and the less-strongly coupled accessory BChl. We also considered cases where only some of the complexes (LH2, LH1, and/or RC) were trimmed. The extent of the trimming (keeping every second BChl) is not critical to our argument: Supplementary Table 4 shows similar results if only every third BChl is kept.

\subsection{Excitonic couplings}

Strong coupling between neighboring BChls in the natural geometry leads to exciton delocalization~\cite{Leupold1996,Pullerits1996,Monshouwer1997,Oijen1999,MayKuhn}. Within each complex (LH2, LH1, or RC), the excitonic states are the eigenstates of a Frenkel Hamiltonian~\cite{MayKuhn} that, in the weak-light, single-exciton regime, reads
\begin{equation}
H=\sum_i E_i \ket{i}\bra{i} + \sum_{i<j} V_{ij}(\ket{i}\bra{j}+\ket{j}\bra{i}),
\label{eq:H}
\end{equation}
where $\ket{i}$ are the site states, $E_i$ are the site energies, and $V_{ij}$ are the couplings. 

The site energies and couplings have been calculated by numerous groups~\cite{Hu1997,Koolhaas1998,Scholes:1999tm,Tretiak2000-2,Linnanto2009}, with the final values varying widely due to different electronic-structure methods and assumptions about the molecular environment. For instance, values between $\SI{238}{cm^{-1}}$ and $\SI{806}{cm^{-1}}$ have been used for the coupling between the $1\upalpha$ and $1\upbeta$ BChls in LH2~\cite{Tretiak2000-2}. To confirm that our conclusions are general and not sensitive to the details of the models, we carried out the entire study using two sets of simulation parameters, which we call $S$ and $R$ because they are inspired by the approaches taken in Schulten's and Renger's groups (Table~\ref{Parameters}). Some differences between the two approaches are worth noting.

Schulten obtains nearest-neighbor couplings from electronic-structure theory (Table~\ref{Parameters}) and more distant ones using the point-dipole approximation (PDA)~\cite{Hu1997,Damjanovic2000,Ritz2001,Sener2007,Strumpfer2009,Sener:2010ct,Strumpfer2012a,Strumpfer2012b}. Although the PDA is accurate at large separations, it can fail even at moderate separations of tens of angstroms, enough to encompass many non-nearest-neighbors~\cite{Howard2004,Fraehmcke:2006ch,Munoz-Losa2009,Kenny2015}. By contrast, Renger's TrEsp method~\cite{Madjet2006} reproduces quantum-chemical couplings with an accuracy better than the unavoidable errors in almost all cases~\cite{Kenny2015}, which is why we use it for all couplings in $R$, except for the special pair, which is too close for TrEsp~\cite{Madjet2009}.

\begin{table}[t]
	\vspace{2mm} 
        \begin{tabular}{lllll}
            \toprule
             & & & \multicolumn{2}{c}{Parameter set} \\
            \cmidrule{4-5}
             & & & \quad$S$ & \quad$R$ \\
            \midrule
            Site   		& LH2	& $E_\upalpha$					& $\num{12458}$~\cite{Strumpfer2009}	& \num{12078}\footnote{Chosen to align brightest state with absorption maximum.} \\
            energies	& 		& $E_\upbeta$						& $\num{12654}$~\cite{Strumpfer2009}	& \num{12274}\footnotemark[1] \\
            (cm$^{-1}$)	& LH1	& $E_\mathrm{LH1}$				& $\num{12121}$~\cite{Damjanovic2000}  & \num{11701}\footnotemark[1] \\
            		  	& RC	& $E_\mathrm{P_1}$	                         & $\num{12180}$~\cite{Strumpfer2012a} 	& \num{11995}\footnotemark[1] \\
			        & 	        & $E_\mathrm{P_2}$                          	& $\num{12080}$~\cite{Strumpfer2012a} 	& \num{11995}\footnotemark[1] \\
            			& 		& $E_\mathrm{B_1}$				& $\num{12500}$~\cite{Strumpfer2012a} 	& \num{12473}\footnotemark[1] \\
            			& 		& $E_\mathrm{B_2}$				& $\num{12530}$~\cite{Strumpfer2012a} 	& \num{12473}\footnotemark[1] \\
            \midrule
            Nearest-   	& LH2	& $V_{1\upalpha 1\upbeta}$			& $363$~\cite{Sener2007}			& ---\footnote{Computed in the same manner as the distant couplings.} \\
            neighbor	& 		& $V_{1\upbeta 2\upalpha}$			& $320$~\cite{Sener2007}			& ---\footnotemark[2] \\
            couplings	& LH1	& alternating\footnote{In $S$, we follow Ref.~\cite{Sener2007} in using alternating couplings even for the non-circular model of LH1.}		& 300, 233~\cite{Sener2007}			& ---\footnotemark[2] \\
          (cm$^{-1}$)	& RC	& $V_\mathrm{P_1P_2}$				& $500$~\cite{Strumpfer2012a}  		& 418 \cite{Madjet2009} \\
            			& 		& $V_\mathrm{P_1 B_1},V_\mathrm{P_2 B_2}$	& $-50$~\cite{Strumpfer2012a}	& ---\footnotemark[2]  \\
            			& 		& $V_\mathrm{P_1 B_2},V_\mathrm{P_2 B_1}$	& $-60$~\cite{Strumpfer2012a} & ---\footnotemark[2]  \\
            \midrule
            More 		&		& Method							& point dipole						& TrEsp\footnote{Using the transition charges in \cite{Madjet2006}.} \\
            distant		&		& $\varepsilon_r$					& 2\footnote{Corresponding to $C=\SI{348000}{\AA^3.cm^{-1}}$ in~\cite{Sener2002}.}	& 1.25\footnote{Corresponding to $f=0.8$ in \cite{Renger2009}.} \\
            couplings	&		& $\mu$ (D)						& 11.75\footnotemark[5]				& 6.1 \cite{Renger2009} \\
	\bottomrule
        \end{tabular}
    \caption{Site energies and couplings. Abbreviations: $\upalpha$ and $\upbeta$: alternating BChls in the B850 unit of LH2; P$_1$, P$_2$: special pair BChls in RC; B$_1$, B$_2$: accessory BChls in RC.}
    \label{Parameters}
\end{table}

The largest differences between $S$ and $R$ are in the relative permittivity $\varepsilon_r$ and the transition dipole moment $\mu$ of each BChl, both of which enter coupling calculations (whether PDA or TrEsp) through the ratio $C=\mu^2/4\pi\varepsilon_0\varepsilon_r$. Schulten sets $C$ so that the PDA reproduces the quantum-chemical coupling (in vacuum) between a particular pair of BChls~\cite{Hu1997,Sener2002,Sener2002a}. This can lead to errors because the long-range couplings now depend on the quantum-chemical method used for the short-range coupling between the calibration pair, as well as on the choice of that pair. Consequently, published values of $C$ have included \num{116000}~\cite{Sener2002a}, \num{146798}~\cite{Strumpfer2012a}, \num{170342}~\cite{Ritz2001}, \num{348000}~\cite{Sener2002}, and $\SI{519310}{\AA^3.cm^{-1}}$~\cite{Hu1997}. We use $\varepsilon_r=2$ and $\mu=\SI{11.75}{D}$, corresponding to $C=\SI{348000}{\AA^3.cm^{-1}}$~\cite{Sener2002,Sener2007,Strumpfer2009}, while noting the unrealistically large dipole moment.

Renger has argued that coupling calculations should use the measured value of $\mu$ \cite{Renger2009,Renger2013}, estimated at \SI{6.1}{D} in vacuum and \SI{7.3}{D} in a medium with $\varepsilon_r=2$ (the effective permittivity of protein complexes~\cite{Renger2009b})~\cite{Knox2003}. Calculations with molecules in cavities surrounded by the dielectric lead to an effective $\varepsilon_r$ from 1.25 to 1.67 \cite{Renger2013}. The recommended combination of $\mu=\SI{6.1}{D}$ and $\varepsilon_r=1.25$~\cite{Renger2009} corresponds to $C=\SI{149000}{\AA^3.cm^{-1}}$. (Alternatively, forgoing the cavity model and using $\varepsilon_r=2$ and $\mu=\SI{7.3}{D}$ gives a similar value, $C=\SI{133000}{\AA^3.cm^{-1}}$.) The large discrepancy between $S$ and $R$ on the value of $C$ has a substantial influence on inter-complex FRET rates, which are proportional to $C^2$.

For the $S$ parameter set, site energies are taken from Schulten's papers (see Table~\ref{Parameters}), and $R$ energies are chosen so that the brightest states of each complex is aligned with the absorption maximum. 

We neglect disorder in site energies, because its effect on the inter-complex transfer rates is small (about 20\% for LH2\,$\to$\,LH2~\cite{Strumpfer2009}) compared to that caused by differences between $S$ and $R$ parameters. The resulting difference in the efficiency would be even less, as is the case for the cyanobacterial photosystem I, where disorder changes the efficiency by about 1\% \cite{Sener2002a}.

\subsection{Energy transfer rates}

The two geometries---natural and trimmed---differ in the inter-pigment couplings, which affects the excitonic states. In the T geometry, we assume that the coupling between the pigments is much weaker than the coupling of the pigments and their environment. In that case, excitons can be thought of as localized on individual sites, and EET is described using F\"orster's theory of resonant energy transfer (FRET)~\cite{MayKuhn}. By contrast, in the N geometry, we assume that the inter-pigment coupling is much stronger than the pigment-bath coupling, leading to delocalized excitons being the better theoretical description. For every aggregate (i.e., each LH2, LH1, and RC), we construct the excitonic states $\ket{\psi_i}$---the eigenstates of that aggregate's Hamiltonian (Eq.~\ref{eq:H})---with dynamics described using Redfield theory~\cite{MayKuhn}. 

The two approaches---full localization and full delocalization---are two ends of a spectrum. The reality---partially localized eigenstates, depending on the details of the system-bath interaction---is somewhere in between. Nevertheless, we focus on these two extremes because they serve to isolate the effect of delocalization on efficiency, which would persist in a diminished form even if the delocalization were only partial.

In all cases, the coupling between different aggregates is weak and is described by FRET. However, delocalization within the aggregates can affect the inter-aggregate FRET rate, potentially leading to supertransfer~\cite{Strek1977,Scholes:2002ie,Lloyd:2010fz}, an effect proposed to be important for the efficiency of purple-bacterial light harvesting~\cite{Strumpfer2012b,Lloyd:2010fz}. 
For example, if the donor contains two pigments and the exciton is localized on either with probability $1/2$, each pigment contributes equally to the total EET rate to the acceptor, $k_\mathrm{loc} \propto \frac{1}{2} |\mu_{\mathrm{D}_1}\mu_\mathrm{A}|^2 +  \frac{1}{2}|\mu_{\mathrm{D}_2}\mu_\mathrm{A}|^2$. But if the exciton is delocalized over the donor sites, the effective transition dipole of the donor is a linear combination of molecular transition dipoles. In the best case, with the exciton in the bright state $(\ket{D_1}+\ket{D_2})/\sqrt{2}$, the transfer rate is doubled, $k_\mathrm{deloc} \propto |\frac{\mu_{\mathrm{D}_1}+\mu_{\mathrm{D}_2}}{\sqrt{2}}\mu_\mathrm{A}|^2=2k_\mathrm{loc}$ if $\mu_\mathrm{D_1}=\mu_\mathrm{D_2}$.

The approximation that the overall transition dipole of the donor eigenstate couples to the acceptor is only valid at large separations. Otherwise, the acceptor sees individual donor sites and not simply a super-molecule, a situation correctly described using generalized FRET (gFRET)~\cite{Mukai1999,Sumi1999,Scholes:2002ie,Scholes2003,Jang2004}, which gives the coupling between a donor state $\psi$ and an acceptor state $\phi$ as 
\begin{equation}
V_{\phi\psi}=\sum_{i,j} c^{\psi}_i c^{\phi}_j V_{ij},
\end{equation}
where $c^{\psi}_i$ ($c^{\phi}_j$) are the site-basis coefficients of the exciton states of the donor (acceptor) and $V_{ij}$ is the coupling between site $i$ of the donor and site $j$ of the acceptor. 
To distinguish ordinary, site-to-site FRET used in the T geometry from the gFRET in the N geometry, we refer to the former as oFRET. 

Once the coupling is known, the FRET rate is \cite{MayKuhn}
\begin{equation}
k_{nm}^{\mathrm{ET}}=\frac{2\pi}{\hbar}|V_{nm}|^2 J_{nm},
\label{eq:FRETrate}
\end{equation}
where the indices $m$ (donor) and $n$ (acceptor) refer to either sites $i$ and $j$ in oFRET or eigenstates $\psi$ and $\phi$ in gFRET. $J_{nm}=\int L_m(E)I_n(E)\,dE$ is the overlap between the normalized fluorescence spectrum $L_m$ of $m$ and the normalized absorption spectrum $I_n$ of $n$. 

The spectra $L_m$ and $I_n$ can be calculated using multichromophoric FRET theory, given a detailed model of each pigment's environment~\cite{Jang2004,Ma:2015kg,Moix:2015ei}. For example, a gFRET calculation of the LH2\,$\to$\,LH2 transfer rate using an Ohmic spectral density \cite{Moix:2015ei} agrees with the same calculation with the exact HEOM method~\cite{Strumpfer2009}. However, transfer rates are sensitive to changes in the spectral density, meaning that substantial errors are introduced by the simplistic assumption of an Ohmic spectral density. Because we doubt that currently available spectral densities are accurate enough to justify computing $L_m$ and $I_n$, we parametrize $J_{nm}$ using experimental data.

We make the simplest possible choice, taking both $L_m(E)$ and $I_m(E)$ to be normalized Gaussian functions centered at $E_m$, with the standard deviation $\sigma=\SI{250}{cm^{-1}}$ chosen to reproduce the width of the Q$_y$ band of BChl\;\textit{a} in solution~\cite{Blankenship2014}. With Gaussian spectra, $J_{nm}$ becomes
\begin{equation}
J_{nm} = e^{-E_{nm}^2/4\sigma^2}/\sqrt{4\pi\sigma^2},
\label{eq:overlap}
\end{equation}
where $E_{nm}$ is the energy difference.
Neglecting the Stokes shift between $L_m$ and $I_m$ would imply $J_{nm}=J_{mn}$, causing Eq.~\ref{eq:FRETrate} to break detailed balance. We correct this by using Eq.~\ref{eq:FRETrate} only for energetically downward transitions ($E_m > E_n$), otherwise taking 
$k_{nm}^{\mathrm{ET}} = k_{mn}^{\mathrm{ET}} e^{-E_{nm}/k_\mathrm{B}T_\mathrm{B}}$, where $T_\mathrm{B}=\SI{300}{K}$ is the ambient temperature.

Transfer rates obtained using this approach agree with those obtained by Schulten's group. For LH2\,$\to$\,LH2 transfer, with the center-to-center distance increased to \SI{85}{\AA}, we obtain a transfer time of \SI{13}{ps}, close to the \SI{9.5}{ps} obtained by Schulten's group using gFRET with their chosen Ohmic spectral density~\cite{Strumpfer2009}. 

\subsection{Relaxation and optical pumping}

A complete model of EET also includes exciton loss (recombination) and creation by optical pumping. 

Excitons can recombine radiatively or non-radiatively. We incorporate radiative relaxation of site/exciton $m$ by adding the rate of spontaneous emission $k^\mathrm{RR}_{gm} = k^\mathrm{RR}_{0} |\mu_{mg}/\mu_0|^2  (E_{mg}/E_0)^3$, where $k^\mathrm{RR}_{0}=(\SI{16.6}{ns})^{-1}$ is the radiative decay rate of BChl in solution~\cite{Monshouwer1997}, $|\mu_{mg}/\mu_0|^2$ is the ratio of the oscillator strength of the transition to that of the single BChl, and $(E_{mg}/E_0)^3$ is the (small) correction to spontaneous emission as the energy of the transition changes with respect to the transition energy of BChl in solution, $E_0=hc/(\SI{770}{nm})$. We include a non-radiative recombination rate of $k^\mathrm{NR}_{gm}=(\SI{1}{ns})^{-1}$ for all $m$~\cite{Sener2007}.

In the natural geometry there is also intra-aggregate relaxation (internal conversion) among the excitonic states, occurring on a sub-picosecond timescale~\cite{vanGrondelle1994}. We incorporate it by assuming, within each aggregate (LH2, LH1, or RC), internal conversion from higher-energy states to lower ones at a rate $k^\mathrm{IC}_{\phi\psi}=(\SI{100}{fs})^{-1}$, while the energetically uphill rates are included by detailed balance. The speed of internal conversion relative to other processes means that each aggregate will be close to a Boltzmann state, as was confirmed for LH2~\cite{Strumpfer2009}.

In natural light, the incoherent light populates system eigenstates and not their superpositions~\cite{Jiang:1991fk,Brumer:2011ty,Mancal2010}, giving rise to a steady state~\cite{Kassal2013,LeonMontiel2014,Pelzer:2014dv}. In principle, these are vibronic eigenstates of the entire light-harvesting apparatus, which raises two considerations. First, eigenstates of the entire apparatus will include superpositions of different complexes. However, inter-complex delocalization is often neglected because it is destroyed by dynamic localization faster than other relevant timescales. This is especially true when the complexes are not energetically resonant and when the system-bath coupling is stronger than the inter-complex coupling, both conditions that apply here. Furthermore, although inter-complex delocalization may increase the absorption of the RC somewhat~\cite{CaycedoSoler:2015tt}, that would modify the efficiency only slightly because most light is absorbed by the antenna complexes and not the RC. Second, the steady-state density matrix of each complex may not be diagonal in the electronic basis due to the system-bath coupling~\cite{Mancal2010,Kassal2013,Olsina:2014vb}. However, our intra-complex Redfield treatment assumes weak system-bath coupling, meaning that off-diagonal elements will be small. We therefore neglect this correction because its influence is likely to be smaller than, say, the difference between $S$ and $R$ parameters. Therefore, we assume that incoherent light populates the sites in the T geometry and the excitonic states of individual aggregates in the N geometry. For site/exciton $m$, the optical pumping rate is $k^\mathrm{OP}_{mg}=k^\mathrm{RR}_{gm}\,n(E_m)$, where $n(E_m)=(e^{E_m/k_\mathrm{B}T_\mathrm{R}}-1)^{-1}$ is the mean photon number at that energy and $T_\mathrm{R}=\SI{5780}{K}$ is the effective black-body temperature of solar radiation. 

The final ingredient is the assumption that excitons in the RC can drive charge separation, at a rate $k^\mathrm{CS}_{gm}=k^\mathrm{CS}=(\SI{3}{ps})^{-1},~m\in \mathrm{RC}$~\cite{Blankenship2014,Sener2007}.

\begin{figure*}[t]
    \includegraphics{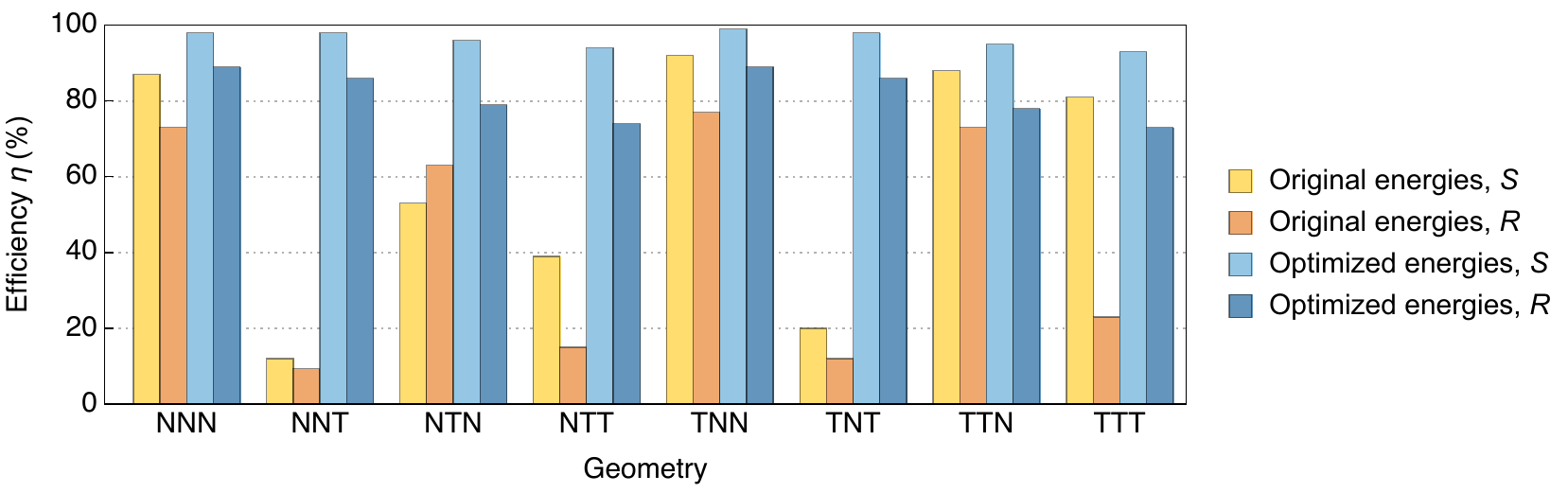}
    \caption{Energy transfer efficiency for various geometries. The three letters indicate whether the natural (N) or trimmed~(T) geometry was used, respectively, for LH2, LH1, and RC. The efficiencies were calculated using the energies in Table~\ref{Parameters} as well as with energies optimized for maximum efficiency, and using both parameter sets $S$ and $R$. The high efficiency in many cases with trimmed complexes indicates that delocalization is not essential for high efficiency. The large improvements when the energies are optimized indicates that energy-level alignment is more important for the efficiency than the effects due to delocalization.}
    \label{fig:barchart}
\end{figure*}

\subsection{Master equation and efficiency}
With all the rates described above, and because there are no coherences in incoherent sunlight, the dynamics of the system can be described using a Pauli master equation,
\begin{equation}
\dot{\mathbf{p}} = K \mathbf{p},
\label{eq:master}
\end{equation}
where $\mathbf{p}$ is the vector containing the populations $p_m$ of all the sites (T geometry) or excitonic states (N geometry), along with the population $p_g$ of the ground state, while the rate matrix $K$ includes all the rates listed above (and summarized in Fig.~\ref{fig:rates}c--d), 
\begin{align}
K_{nm} &= k_{nm}^{\mathrm{ET}} + k_{nm}^{\mathrm{RR}} + k_{nm}^{\mathrm{NR}} + k_{nm}^{\mathrm{IC}} + k_{nm}^{\mathrm{OP}} + k_{nm}^{\mathrm{CS}}  \\
		&  \omit \hfill  (for $n\ne m$),\notag \\
K_{mm} &= -\! \sum_{n\ne m} K_{nm}.
\end{align}

We define the efficiency as the quantum yield of charge separation, i.e., the probability that a photon absorbed by any of the complexes eventually drives charge separation in the RC. Because energy can be lost along the way, this is not a thermodynamic efficiency.

Because incoherent excitation is stationary, the molecular ensemble will be at steady state, 
$\mathbf{p}^\mathrm{SS}$~\cite{Kassal2013,LeonMontiel2014,Pelzer:2014dv}. Since $\dot{\mathbf{p}}^\mathrm{SS}=K\mathbf{p}^\mathrm{SS}=0$, $\mathbf{p}^\mathrm{SS}$ can be easily found as the unique eigenvector of $K$ with eigenvalue zero. The efficiency is then the rate of charge separation in the RC divided by the rate at which excitons are created,
\begin{equation}
\eta = \frac{k^\mathrm{CS} \sum_{m\in \mathrm{RC}} \, p_m^\mathrm{SS}}{p_g^\mathrm{SS}\,\sum_m k^\mathrm{OP}_{mg}},
\label{eq:efficiency}
\end{equation}
where $p^\mathrm{SS}_g$ is the steady-state population of the ground state. The $k^\mathrm{OP}$ in the denominator ensures $\eta$ is intensity-independent as long as the light is weak and $p^\mathrm{SS}_g\approx 1$.

Eq.~\ref{eq:efficiency} can be compared to the efficiency used by Schulten's group, $\eta=-k^\mathrm{CS} (\mathbf{p}^\mathrm{RC})^\top K^{-1} \mathbf{p}^\mathrm{I}$, where $\mathbf{p}^\mathrm{RC}$ is a uniform distribution over the RC sites and $\mathbf{p}^\mathrm{I}$ is the initial distribution~\cite{Ritz2001,Sener2007}. This definition is not conceptually suited to steady-state light harvesting, where there is no ``initial'' state, although the two approaches are equivalent if a suitable ``initial'' state is chosen \cite{jesenko2013}. Because the initial state can influence the efficiency substantially~\cite{LeonMontiel2014}, it is important to model the light absorption and not assume an initially localized state~\cite{Sener2007,Strumpfer2009} or a uniform distribution over many~\cite{Sener2002a}.

\section{Results and Discussion}

\subsection{Original model}

The overall efficiencies of the various cases are summarized in Fig.~\ref{fig:barchart}, with details, including inter-complex transfer rates, in Supplementary Table 1. The results obtained using the $S$ and $R$ parameters can be quite different due to the weaker long-range coupling in $R$, confirming our concern about the difficulty of precise calculations and the need to focus on general trends. 

Figure~\ref{fig:barchart} shows that delocalization is not necessary for high efficiency. In particular, in the $S$ parameters (but not $R$---see below) the trimmed geometry TTT is as efficient as the natural geometry NNN. In addition, delocalization in LH2 always diminishes the efficiency: with both $S$ and $R$ parameters, $\eta_\mathrm{TXY} > \eta_\mathrm{NXY}$ for all $\mathrm{X,Y\in\{N,T\}}$, disproving the hypothesis that delocalization in LH2 benefits the efficiency through supertransfer. 
By contrast, delocalization in the RC is always beneficial, while in LH1 delocalization benefits the efficiency if and only if there is delocalization in the RC. 

Whether delocalization is beneficial or deleterious can be explained using Eq.~\ref{eq:FRETrate}. Delocalization affects the EET rate by altering both $|V_{nm}|^2$ (potentially yielding supertransfer) and the excitonic energy gaps (through the Davydov splittings). Indeed, the spectral overlap is exponentially sensitive to excitonic splittings (Eq.~\ref{eq:overlap}), unlike supertransfer, which is at best an improvement by a constant factor equal to the number of sites.

\begin{figure*}[t]
    \centering
    \includegraphics{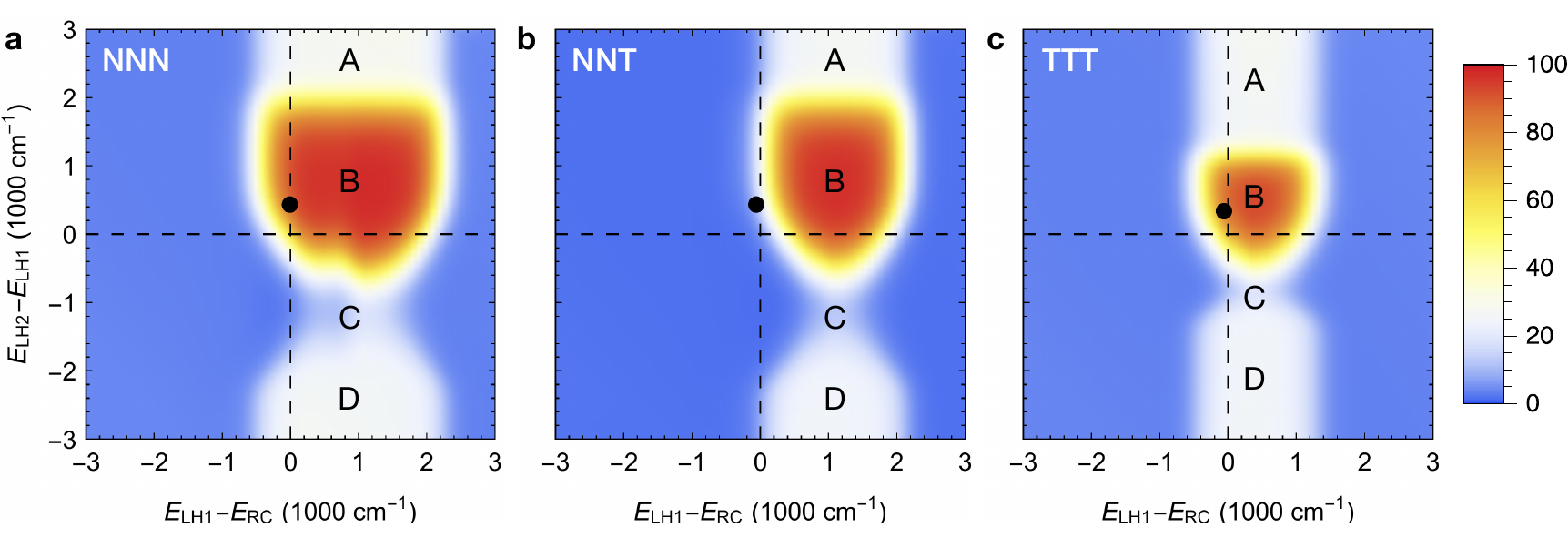}
    \caption{Efficiency of energy transfer as a function of the energy gaps $E_\mathrm{LH1}-E_\mathrm{RC}$ and $E_\mathrm{LH2}-E_\mathrm{LH1}$, for the cases \textbf{(a)} NNN, \textbf{(b)} NNT, and \textbf{(c)} TTT, all in the $S$ parameter set. The efficiency is sensitive to the site energies, with the maximum occurring when there is a clear energy funnel, $E_\mathrm{LH2} > E_\mathrm{LH1} > E_\mathrm{RC}$, provided that the differences in energy are not so large as to inhibit EET through a diminished spectral overlap. The original energy gaps are indicated by the black dots, and the features A--D are discussed in the text. The corresponding plots for the $R$ parameter set are Supplementary Fig.~2.
    }
    \label{fig:optimisation}
\end{figure*}

Delocalization in LH2 is deleterious because it weakens the energy funnel into LH1. LH2 sites are higher in energy than LH1 sites, ensuring that backward transfer LH1\,$\to$\,LH2 is suppressed. If the excitons are delocalized, the bright state of LH2 is lowered in energy by the Davydov splitting, bringing it closer to LH1. Although this encourages forward transfer, it increases backward transfer even more because of detailed balance. Furthermore, because there are 10 LH2s and only one LH1, there are 10 times as many pathways back to LH2. Consequently, the highest efficiencies occur when the forward rate is much greater than the backward rate (see Supplementary Table 1).
In particular, the main pathway between LH2 and LH1 is through their brightest eigenstates, which have a gap of about $\SI{350}{cm^{-1}}$. If LH2 is trimmed, EET occurs through the sites, whose energy is about $\SI{1000}{cm^{-1}}$ above the brightest state of LH1. The increased gap makes backward EET exceedingly slow, while keeping the forward rate adequate for high efficiency. 

The analysis in the RC is analogous. Its site energies are higher than in LH1, which would inhibit efficient transport. Including the excitonic couplings splits the levels, brining the lower one closer to LH1 in energy and increasing the rate of forward EET~\cite{Strumpfer2012b}. The efficiency is worst when LH1 is delocalized and RC is not (NNT and TNT), because that maximizes the energy difference between LH1 and the RC.

Of course, the spectral overlap does not offer the complete explanation. Although the smallest energy gap between LH1 and RC occurs when both complexes are trimmed, the highest LH1\,$\to$\,RC rate is when both are delocalized. This is due both to the brightness of the exciton states in both complexes (supertransfer) and the larger number of pathways toward the RC when there are four BChls as opposed to two. The importance of the number of pathways (i.e., entropic driving) is further illustrated in Supplementary Table 3.

\subsection{Energy optimization}

The preceding discussion suggests that excitonic couplings affect EET rates more through energy-level shifts than through supertransfer. If so, the poor performance in cases such as NNT and TNT should be correctable by adjusting the site energies.  

We tested this hypothesis by repeating the simulations with site energies as free parameters. 
BChl site energies vary widely in photosynthetic complexes: single substitutions on the protein backbone can significantly affect the  energy, especially if the residue axially ligates the magnesium or if is charged and thus modifies the electrostatic environment~\cite{Renger2013}. It is therefore plausible that natural selection could have modified the site energies (within limits, of course) if it increased fitness.

As Fig.~\ref{fig:barchart} shows, choosing the optimal value for the site energies of LH1 and LH2 (relative to the average special-pair energy $E_\mathrm{RC}$) dramatically enhances the efficiency in all cases. In the $S$ parameters, the optimized efficiency is always above $93\%$. The $R$ parameters do not perform as well  due to weaker long-range couplings, but the optimal efficiency is nevertheless always above $73\%$, i.e., the same or greater than the natural-case efficiency of $73\%$. For details, see Supplementary Table~2.

The behaviour of the efficiency with changes in the site energies $E_\mathrm{LH1}$ and $E_\mathrm{LH2}\equiv\frac12 (E_\upalpha+E_\upbeta)$ is shown in Fig.~\ref{fig:optimisation}. Most features of the plots can be understood as consequences of energy funnelling. For example, the efficiency is negligible unless LH1 is higher in energy than the RC or only slightly (a few $\sigma$) lower, but not so high that their spectra no longer overlap.

When LH2 is also considered, there are four regions of interest, labeled A--D in Fig.~\ref{fig:optimisation}, depending on the offset $E_\mathrm{LH2}-E_\mathrm{LH1}$. The peak B occurs when there is a clear energy funnel toward LH1, whereas in the adjacent valley C, LH2 is slightly lower than LH1, so that outward transfer LH1\,$\to$\,LH2 dominates. This problem is compounded by the number of available pathways (i.e., entropy): because there are 10 LH2s in the model, an exciton on LH1 can move to an LH2 in 10 different ways (at the same rate), while an exciton on LH2 has only one pathway to LH1. Therefore, the excitons accumulate in the LH2s, making the transfer to the RC unlikely.

When the offset $E_\mathrm{LH2}-E_\mathrm{LH1}$ is large, whether positive (A) or negative (D), the small spectral overlap between LH1 and LH2 gives a small efficiency. In those cases, only excitons starting out in LH1 or the RC contribute to the efficiency, while those from LH2---by far the largest proportion---cannot leave and are wasted. The efficiency in region A is higher than in region D (and increases further in the unphysical limit $E_\mathrm{LH2}\to\infty$) because there are fewer photons at higher energies, meaning that fewer excitons start out at LH2 and are wasted.

The plots for the three cases in Fig.~\ref{fig:optimisation} (and for all the cases that are not shown) are qualitatively the same, differing only in the positions of the peaks and their widths. The differences in peak positions are caused by the excitonic splittings, while the different widths occur because the couplings allow both lower- and higher-energy excitonic states to act as EET donors and acceptors.

\subsection{Role and evolution of coherence}

Delocalization in LH2 prompted speculation that it plays a functional role, having been selected by evolution because it enhanced light-harvesting efficiency. However, this is not so, since trimming the LH2 always increases the efficiency. Why then is there delocalization? The main evolutionary pressure is probably not on efficiency but on the total number of excitons processed by each RC. Although removing half the BChls might increase the efficiency marginally, it would also halve the absorption cross-section of LH2, decreasing the total exciton flux into the RC. In other words, delocalization in LH2 is a spandrel~\cite{Gould:1997fv}, a byproduct of packing BChls densely to maximize absorption, just as the red color of vertebrate blood is not adaptive, but a byproduct of the oxygen-carrying ability of haemoglobin, which happens to be red. Indeed, a clade of cryptophyte algae underwent a mutation that reduced excitonic delocalization in their antennas~\cite{Harrop:2014io} with no apparent decrease in fitness.

By contrast, the N geometry in the RC always outperforms T because the Davydov splitting brings its energy closer to that of LH1.  However, it would put the cart before the horse to conclude that RC energy splittings are an adaptation to create better energy alignment with LH1. Rather, a strongly coupled special pair is a feature conserved across photosynthetic organisms~\cite{Blankenship2014}, meaning that it arose before the advent of purple bacteria.

Instead, we should ask whether purple bacteria built the antennas around the RC so as to create an effective energy funnel. The answer is yes, as confirmed by the high efficiency in the natural case. Even so, the alignment is not optimal, and Fig.~\ref{fig:barchart} indicates a substantial further increase in efficiency is possible in principle, depending on how much further the site energies could realistically be altered. Nevertheless, the continued survival of the species indicates that the efficiency is probably good enough that an additional increase would not confer a decisive evolutionary advantage.

In particular, having the RC higher in energy than LH1 creates a rate-limiting uphill step at the end of the EET chain. It has been proposed that the uphill transfer is also adaptive, preventing too many excitons arriving at a `closed' RC (one having recently undergone charge separation), thus averting excessive energy dissipation and thermal damage~\cite{Ritz2001,Sener2007,Strumpfer2012a}. However, we doubt that the exciton flux would be large enough to damage the RC, considering that the light-harvesting processes are completed orders of magnitude faster than the mean time between the arrivals of two photons. Therefore, it appears that the final uphill step is, like delocalization, an evolutionary relic in a light-harvesting system that functions well enough, but not optimally.

\section{Conclusions}

EET efficiency in purple bacteria can be largely understood in terms of the energy landscape, with supertransfer playing a minor role. Because FRET rates are exponentially sensitive to energy offsets, small changes in site energies---well within the range found in nature---can often improve the rates more than the maximal effect of supertransfer.  Despite this sensitivity, the purple-bacterial energy funnel is robust, and would yield high efficiencies across a considerable range of site energies. The dominance of energy funnelling is seen using two very different parameter sets, confirming that it is a sturdy conclusion insensitive to the details of the model.

\section*{Acknowledgements}
We thank Vahid Karimipour, Elise Kenny, and Saleh Rahimi-Keshari for valuable discussions.
We were supported by the Australian Research Council through a Discovery Early Career Researcher Award (DE140100433) and the Centres of Excellence for Engineered Quantum Systems (CE110001013) and Quantum Computation and Communication Technology (CE110001027).  


%

\onecolumngrid
\setcounter{table}{0}
\setcounter{figure}{0}
\renewcommand{\figurename}{Supplementary Figure}
\renewcommand{\tablename}{Supplementary Table}
\renewcommand{\tabcolsep}{4.5pt}
\renewcommand{\thetable}{\arabic{table}} 

\section*{\\[5mm]\large{D\lowercase{istinguishing the roles of energy funnelling and delocalization in photosynthetic\\ light harvesting}\\~\\S\lowercase{upplementary Information}}}

\vskip5mm
\begin{figure}[H]
    \centering
    \includegraphics{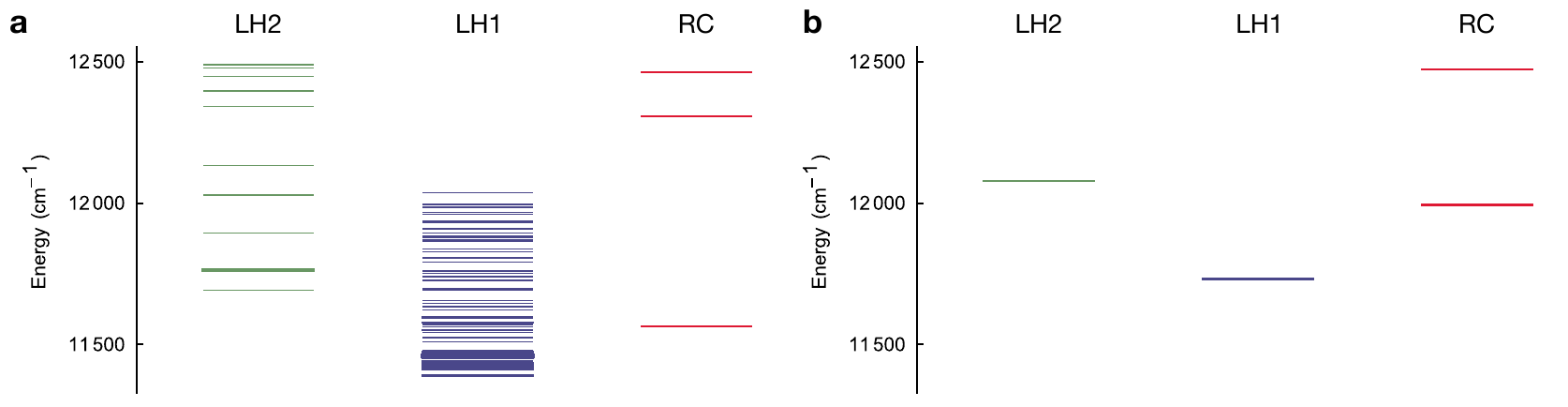}
    \caption{Energy levels of the natural \textbf{(a)} and trimmed \textbf{(b)} complexes in the $R$ parameter set. An important difference with respect to the $S$-parameter levels shown in Fig.~2 is that LH1\,$\to$\,RC transfer is significantly uphill in the trimmed geometry, explaining the low efficiency of the TTT configuration in the $R$ parameter set.}
    \label{fig:energylevels-renger}
\end{figure}

\begin{table}[H]
\centering
\begin{tabular}{cccccSSSSS}
\toprule
& \multicolumn{3}{c}{Geometry} & & \multicolumn{5}{c}{Transfer times (ps)}  \\ 
\cmidrule{2-4} \cmidrule{6-10}
& LH2 & LH1 & RC & {\quad$\eta~(\%)$\quad} & {LH2$\,\to\,$LH2} & {LH2$\,\to\,$LH1} & {LH1$\,\to\,$LH2} & {LH1$\,\to\,$RC} & {RC$\,\to\,$LH1} \\
\midrule
{\;\parbox[t]{3ex}{\multirow{8}{*}{\rotatebox[origin=c]{90}{$S$ parameters}}}}
& N & N & N & 87 & 4.5 & 3.2 & 35 & 25 & 1.9 \\
& N & N & T & 12 & 4.5 & 3.2 & 35 & 3000 & 19 \\
& N & T & N & 53 & 4.5 & 16 & 17 & 72 & 55 \\
& N & T & T & 39 & 4.5 & 16 & 17 & 82 & 5.3\\
& T & N & N & 92 & 9.6 & 12 & 1900 & 25 & 1.9\\
& T & N & T & 20 & 9.6 & 12 & 1900 & 3000 & 19\\
& T & T & N & 88 & 9.6 & 13 & 200 & 73 & 55\\
& T & T & T & 81 & 9.6 & 13 & 200 & 82 & 5.3\\
\midrule
{\;\parbox[t]{3ex}{\multirow{8}{*}{\rotatebox[origin=c]{90}{$R$ parameters}}}}
& N & N & N & 73 & 17 & 17 & 340 & 156 & 7.6 \\
& N & N & T & 9.3 & 17 & 17 & 340 & 6400 & 41 \\
& N & T & N & 63 & 17 & 26 & 171 & 190 & 27 \\
& N & T & T & 15 & 17 & 26 & 172 & 2200 & 42\\
& T & N & N & 77 & 48 & 43 & 2400 & 156 & 7.6\\
& T & N & T & 12 & 48 & 43 & 2400 & 6400 & 42\\
& T & T & N & 73 & 48 & 78 & 1500 & 190 & 27\\
& T & T & T & 23 & 48 & 78 & 1500 & 2200 & 42\\
\bottomrule
\end{tabular}
\caption{Efficiencies ($\eta$) and inter-complex energy transfer times when the site energies are the original energies listed in Table~1.
Our results for the natural geometry NNN in the $S$ parameter set (first line above) are consistent with those of Schulten's group. In particular, our departures from Schulten's approach---treating fewer light-harvesting complexes and using the simplified overlap integral in Eq.~4---are justified because they do not substantially affect the results. Our efficiency and our transfer times for LH1\,$\to$\,RC and RC\,$\to$\,LH1 are similar to Schulten's values of, respectively, 78--91\% (depending on vesicle structure), \SI{20}{ps}, and \SI{1.4}{ps}~\cite{Sener2007}. Our transfer times involving LH2 are shorter because we assumed a tighter packing based on AFM measurements (compare with Schulten's \SI{10}{ps} for both LH2\,$\to$\,LH2 and LH2\,$\to$\,LH1~\cite{Sener2007}). As noted in the text, this difference largely disappears if the distances are the same.
The $S$ parameters tend to agree with experiment somewhat better than the $R$ parameters, although not completely. For example, $S$ parameters get the LH1\,$\to$\,RC transfer time approximately right (estimated at $\SI{20}{ps}$ at room temperature~\cite{vanGrondelle1994,Bergstrom1989}), but the reverse rate (7--9~ps \cite{Timpmann1993}) is much better captured by the $R$ parameters. Measured transfer times involving LH2 ($\SI{5}{ps}$ for LH2\,$\to$\,LH2~\cite{Agarwal2002} and $\SI{3.3}{ps}$ for LH2\,$\to$\,LH1~\cite{Hess1995}) also agree with those predicted from the $S$ parameters at the separations we used.   }
\label{tab:NaturalSiteEnergies}
\end{table}

\begin{table}[H]
\centering
\begin{tabular}{cccccSSSSS}
\toprule
& \multicolumn{3}{c}{Geometry} & & \multicolumn{5}{c}{Transfer times (ps)}  \\ 
\cmidrule{2-4} \cmidrule{6-10}
& LH2 & LH1 & RC & {\quad$\eta~(\%)$\quad} & {LH2$\,\to\,$LH2} & {LH2$\,\to\,$LH1} & {LH1$\,\to\,$LH2} & {LH1$\,\to\,$RC} & {RC$\,\to\,$LH1} \\
\midrule
{\;\parbox[t]{3ex}{\multirow{8}{*}{\rotatebox[origin=c]{90}{$S$ parameters}}}}
& N & N & N & 98 & 4.5 & 4.5 & 140 & 4.9 & 53 \\
& N & N & T & 98 & 4.5 & 5.2 & 220 & 7.4 & 9.6 \\
& N & T & N & 96 & 4.5 & 10 & 390 & 23 & 80 \\
& N & T & T & 94 & 4.5 & 12 & 640 & 32 & 13\\
& T & N & N & 99 & 9.7 & 6.4 & 240 & 4.9 & 53\\
& T & N & T & 98 & 9.7 & 7.4 & 410 & 7.4 & 9.6\\
& T & T & N & 95 & 9.7 & 18 & 500 & 23 & 89\\
& T & T & T & 93 & 9.7 & 21 & 700 & 32 & 13\\
\midrule
{\;\parbox[t]{3ex}{\multirow{8}{*}{\rotatebox[origin=c]{90}{$R$ parameters}}}}
& N & N & N & 89 & 18 & 28 & 1500 & 70 & 480 \\
& N & N & T & 86 & 18 & 32 & 2100 & 96 & 30 \\
& N & T & N & 79 & 20 & 57 & 2000 & 150 & 1100 \\
& N & T & T & 74 & 17 & 67 & 2800 & 190 & 37\\
& T & N & N & 89 & 48 & 42 & 2200 & 70 & 480\\
& T & N & T & 86 & 48 & 45 & 2600 & 96 & 30\\
& T & T & N & 78 & 48 & 98 & 2500 & 150 & 1100\\
& T & T & T & 73 & 48 & 110 & 3500 & 190 & 37\\
\bottomrule
\end{tabular}
\caption{Efficiencies ($\eta$) and energy transfer times when the site energies are optimised for each geometry.}
\label{tab:OptimalSiteEnergies}
\end{table}

\vskip-.5cm
\begin{figure}[H]
    \centering
    \includegraphics{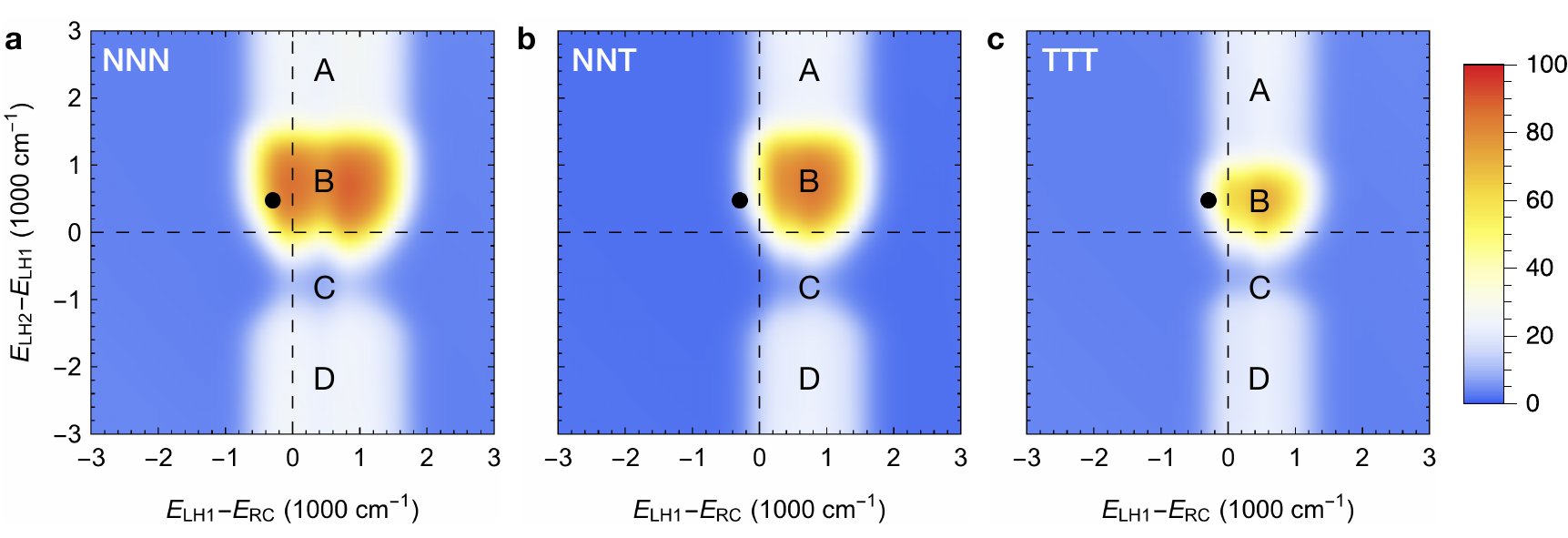}
    \caption{As Fig.~4, but for the $R$ parameter set.}
    \label{fig:optimisationR}
\end{figure}

\begin{table}[H]
\centering
\begin{tabular}{cccccSSSSS}
\toprule
& \multicolumn{3}{c}{Geometry} & & \multicolumn{5}{c}{Transfer times (ps)}  \\ 
\cmidrule{2-4} \cmidrule{6-10}
& LH2 & LH1 & RC & {\quad$\eta~(\%)$\quad} & {LH2$\,\to\,$LH2} & {LH2$\,\to\,$LH1} & {LH1$\,\to\,$LH2} & {LH1$\,\to\,$RC} & {RC$\,\to\,$LH1} \\
\midrule
{\;\parbox[t]{3ex}{\multirow{4}{*}{\rotatebox[origin=c]{90}{$S$ params}}}}
& N & N & $\mathrm{T}^+$ & 28 & 4.5 & 3.2 & 35 & 1100 & 16\\
& N & T & $\mathrm{T}^+$ & 58 & 4.5 & 16 & 17 & 40 & 6.5\\
& T & N & $\mathrm{T}^+$ & 41 & 9.6 & 12 & 1900 & 1100 & 16\\
& T & T & $\mathrm{T}^+$ & 90 & 9.6 & 13 & 200 & 40 & 6.5\\
\midrule
{\;\parbox[t]{3ex}{\multirow{4}{*}{\rotatebox[origin=c]{90}{$R$ params}}}}
& N & N & $\mathrm{T}^+$ & 16 & 17 & 17 & 341 & 3500 & 45\\
& N & T & $\mathrm{T}^+$ & 24 & 17 & 26 & 172 & 1200 & 46\\
& T & N & $\mathrm{T}^+$ & 21 & 48 & 43 & 2400 & 3500 & 45\\
& T & T & $\mathrm{T}^+$ & 35 & 48 & 78 & 1500 & 1200 & 46\\
\bottomrule
\end{tabular}
\caption{Effect of the number of pathways on the efficiency. 
As noted in the text, having more pathways going away from the RC than towards it implies that entropy drives charges away from the RC. To illustrate this further, we increased the number of BChls in the trimmed RC by replacing the ones that were discarded, indicated as T$^+$. In other words, we used ordinary FRET on the natural RC geometry, even if that is not the correct description of EET. It can be seen that doubling the number of BChls in each RC roughly doubles the transfer rate LH1\,$\to$\,RC with respect to the values in Supplementary Table~\ref{tab:NaturalSiteEnergies}. This results in an appreciable increase in the efficiency, averaging $75\%$ for the $S$ parameters and $65\%$ for the $R$ parameters (relative to Supplementary Table~\ref{tab:NaturalSiteEnergies}).}
\label{tab:MorePathways}
\end{table}

\begin{table}[H]
\centering
\begin{tabular}{ccccSSSSS}
\toprule
\multicolumn{3}{c}{Geometry} &  & \multicolumn{2}{c}{$S$ parameters} & & \multicolumn{2}{c}{$R$ parameters}  \\ 
\cmidrule{1-3} \cmidrule{5-6}  \cmidrule{8-9}
LH2 & LH1 & RC & & {$\eta_\text{Orig}~(\%)$} &  {$\eta_\text{Opt}~(\%)$} & & {$\eta_\text{Orig}~(\%)$} &  {$\eta_\text{Opt}~(\%)$} \\
\midrule
 N & T$^3$ & N & & 44 & 95 & & 57 & 76 \\
 N & T$^3$ & T & & 34 & 93 & & 12 & 70 \\
 T$^3$ & N & N & & 92 & 98 & & 78 & 90 \\
 T$^3$ & N & T & & 21 & 98 & & 13 & 86 \\
 T$^3$ & T$^3$ & N & & 89 & 95 & & 72 & 76 \\
 T$^3$ & T$^3$ & T & & 84 & 93 & & 25 & 70 \\
\bottomrule
\end{tabular}
\caption{EET efficiencies when the LH2 and/or LH1 aggregates are trimmed more aggressively. T$^3$ denotes aggregates in which only every third BChl is kept and the results are shown using both the original site energies and the optimal ones. The efficiencies are similar to those obtained with the every-second trimming described in the text, indicating that our conclusions are not sensitive to the extent of trimming.}
\label{tab:FurtherTrimming}
\end{table}

\end{document}